\documentstyle[12pt]{article}
\newcommand{\be}{\begin{equation}}   \newcommand{\ee}{\end{equation}}
\newcommand{\bear}{\begin{eqnarray}}
\newcommand{\eear}{\end{eqnarray}}
\newcommand{\ba}{\begin{array}}      \newcommand{\ea}{\end{array}}

\begin{document}
\par \vskip .05in
\begin{titlepage}
\begin{flushright}
FERMI--PUB--00/047--T\\ 
EFI--2000--6; hep-ph/0002230\\
Feb. 22, 2000 \\
Submitted to {\em Phys. Rev. D}
\end{flushright}
\vfill
\begin{center}
{ 
\Large \bf 
The Diffractive Quantum Limits of \\
\vskip .1in Particle Colliders 
}
\end{center}
\par \vskip .1in \noindent
\begin{center}
{\bf \large   Christopher T.~Hill$^{1,2}$ 
}
  \par \vskip .1in \noindent
{$^1$Fermi National Accelerator Laboratory\\
 P.O. Box 500, Batavia, Illinois, 60510}
  \par \vskip .05in \noindent
{$^2$The University of Chicago\\
Enrico Fermi Institute, Chicago, Illinois, 60637}
\end{center}     
\par \vskip .05in
\begin{quote}
   {\normalsize
Quantum Mechanics places limits on 
achievable transverse beam spot sizes of particle
accelerators. We estimate 
this limit for a linear collider to be
$\Delta x \geq \hbar c f/E\delta_0 $ where $f$ is the final focal length, 
$E$ the
beam energy, 
and $\delta_0$ the intrinsic transverse Gaussian width of 
the electron wave-function. $\delta_0$ is determined in the
phase space damping rings, and we find
$\delta_0 \approx \sqrt{ \hbar c /eB}$ where $B$ is the typical wiggler
magnetic field strength in this system. For the
NLC $\delta_0 \sim 25$nm, and $\Delta x \sim {\cal{O}}(0.06)$ nm, 
about two orders of magnitude smaller than the design goal.  
We can recover a crude estimate of the classical
result when we include radiative relaxation effects.
We also consider a synchrotron and 
obtain $\Delta x \geq \sqrt{\hbar c f/E} \sim {\cal{O}}(1.0)$ nm,
We discuss formulation
of quantum beam optics relevant to these issues.
}
\end{quote}
\vfill
\end{titlepage}

\baselineskip=18pt
\pagestyle{plain}
\setcounter{page}{1}

\relax
\pagestyle{plain}

\section{Introduction}

Particle accelerators are 
designed and built, based essentially
upon the classical
theory of point charges interacting with
electromagnetism.  Nevertheless, particles are
described by wave-functions, and diffractive
limits must exist as to how well
they can be localized in
a given optical 
apparatus. 
The first quantum mechanical effects to arise in a potentially 
limiting way might be expected to be diffractive in nature. 
In this paper we take a first look at the problem of estimating
the quantum diffractive limits of accelerators.
We begin with an important system, an NLC-class machine.
We are inspired to consider this because the desired goals for
the NLC beam spot size are ambitious. To achieve
the desired luminosity requires a $\sim 5$ nm beam spot in one
transverse dimension, (the vertical or $y$ direction
in the NLC reports
\cite{nlc}).  We will find that this criterion is about two
orders of magnitude above the quantum limit. Indeed, we will describe
how to estimate the classical design result for the
beam spot size itself 
from quantum mechanics, obtaining rough agreement with the
NLC specifications. 

We obtain a conceptually simple result. The diffractive
limit on the beam spot size in the $x_\perp $ direction is given by
the Rayleigh formula for a (massless) wave of energy $E$ which
has passed through an effective
``aperature'' $\delta_0$ and focused over
a focal length $\sim f$. That is:
\be
\Delta x \geq \frac{\hbar c f}{E \delta_0}
\ee
where $f$ is the final focal length, $E$ the beam energy
(the result varies somewhat in a compound
lens system, see Section 3).
We emphasize that the ``aperature'' $\delta_0$ is {\em not} a
mechanical aperature, e.g., it is not the beam pipe size.
$\delta_0$ is actually 
the initial state Gaussian width of the 
transverse quantum wave-function as
it enters the linac upstream 
from the damping rings, where the wave-function
has been prepared (we assume an ``ideal linac,''
in which there is negligible further synchrotron radiation
downstream;
this is not necessarily a good approximation, and corrections
to the effective $\delta_0$ are expected).
The initial state can be considered 
to be an ensemble of particles, each
in simple harmonic oscillator (SHO) transverse
wave-functions, where the 
Gaussian envelope (groundstate) width is
determined by the damping ring wiggler system.
This is given by:
\be
\delta_0 = \sqrt{\frac{\hbar c}{eB}}
\ee
where $B$ is the typical magnetic field in the damping system,
of order $1$ Tesla.
Taking $f=2$m, $E=250$ GeV, $B=1$ Tesla, yields $\delta_0 \approx 25.7$ nm,
and thus,
$\Delta x > 0.062$ nm
as a diffracive limit.  
Hence, the NLC would appear to be safely above the quantum limit
by about two orders of magnitude.
We remark, however, that this is the extremal lower limit which
saturates the Heisenberg uncertainty relationship, and
holds in our idealized limit.  

In
general the individual particle 
initial state is an excited SHO transverse wave-function of 
average principle
quantum number $\bar{n}$.  This increases the expected
diffractive spot size to $\Delta x \approx \sqrt{\bar{n}}{f}/{E \delta_0}$.
In fact, since $\bar{n} \propto 1/\hbar$  it is easily
seen that this result is independent
of $\hbar$, and therefore should be equivalent to a 
classical derivation
of the beam spot size.   
$\bar{n}$ can be crudely estimated from radiation
relaxation following the original
arguments of Sands \cite{sands}, and others.
This yields a result of $\Delta x \approx 2$ nm, 
roughly consistent
with the NLC design report calculations for
the vertical beam spot \cite{nlc}.

The subject of quantum
beam dynamics for particle
accelerators is fairly novel \cite{chen}.
Gaussian optics is a preferred formalism for
tackling the problem considered here. 
Presently we construct a transverse Gaussian wave-packet,
with a longitudinal plane wave structure, and propagate it
through an optical system. 
Gaussians
extremalize the Heisenberg uncertainty relation, and they 
are also the groundstate solutions in continuous
linear focusing channels, and e.g., magnetic
lenses, wigglers, to a reasonable approximation,
etc., and can be approached by synchrotron
radiation relaxation \cite{sands}, \cite{sokolov}, \cite{huang}, \cite{huang2}. 
Remarkably, Gaussian transverse wave-functions, which
solve the quantum Schroedinger equation for 
propagation through the optical system,
(neglecting synchrotron radiation), are controlled
entirely by the classical lens matrices of the system.
While Gaussian optics is a standard formalism
in treating electron microscopy
\cite{glaser}, \cite{arnaud}, \cite{fock}, to our knowledge, the 
behavior of a quantum Gaussian beam in a synchrotron
has not been previously formulated, and we
will indicate the self-replicating solution to a synchrotron
by an application of lens matrix methods. 

First consider the problem of a
relativistic electron wave-function passing
though a  lens. Spin is an inessential complication
\cite{jagan}, \cite{khan}, 
so we can use the Klein-Gordon (KG) equation.
Assume for simplicity
that there is only one  spatial transverse dimension, $x_\perp $, 
and let $z$ be the longitudinal spatial
dimension. In the 
KG equation we include a transverse
simple harmonic oscillator  (SHO) potential term
which is dependent upon $z$ (For the analysis, we 
set $\hbar=c=1$):
\be
\partial^2 \phi +m^2\phi +  \tilde{K}(z) x_\perp^2 \phi = 0
\ee
Then,
with $\phi =\exp[-i(Et-p_zz)]\hat{\phi}$
and $E^2=p_z^2 + m^2$, the KG
equation becomes the transverse Schoedinger equation:
\be
\label{eq0}
i \frac{\partial}{\partial z}\hat{\phi} +\frac{1}{2E}(\vec{\nabla }_\perp 
)^2\hat{\phi} - \frac{K(z) x^2}{2} \hat{\phi} =0
\ee
where: 
\be
K \equiv \tilde{K}/E.
\ee
This is a standard construction in optics \cite{arnaud},
and $\hat{\phi}(z)$ has the conventional interpretation with $z$
replacing time.  
The parameter $K(z)$ 
is $z$-dependent, corresponding to the
finite longitudinal structure of the lens system.  
For
a single thin lens we take $K(0)=0$ for $z< -\delta z $ and $z > 0$,
and $K=K_0 $ for $-\delta z \leq z \leq 0$.

Let us now postulate
a Gaussian form for the wave-function centered at 
the transverse position $x_\perp$,
carrying a transverse momentum $p_\perp$:
\be
\label{eq4}
\hat{\phi} = \exp\left( -\frac{1}{2} A(z)(x-x_\perp)^2 +ip_\perp x + C  \right)
\ee
In this expression, $A(z)$ is the complex Gaussian
kernel, $x_\perp$ and $p_\perp$ are real,
and $C$ simply parameterizes the
overall normalization. Hence, 
the Gaussian wave-function has four real parameters.
After substituting this anzatz, the Schroedinger equation, eq.(\ref{eq0}),
yields the following equations of motion for the width:
\be
\label{eq1}
i\frac{\partial A}{\partial z} = \frac{ A^2}{E} - K(z)
\ee
and for $x_\perp$ and 
$p_\perp$ we obtain 
the classical Hamilton equations:
\be
\label{eq2}
\frac{\partial p_\perp}{\partial z} = - K x_\perp;
\qquad
\frac{\partial x_\perp}{\partial z} = \frac{p_\perp}{E};
\ee
Note that the centroid $x_\perp$ and 
centroid momentum $p_\perp$ motions
are decoupled from that of the
Gaussian kernel $A(z)$ and vice versa. 
(anharmonic effects would generally couple these quantities).
Moreover, the boundary conditions on $A(z)$
and of the centroid $x_\perp$ and 
centroid momentum $p_\perp$ are independent .
Note that the last equation is just the ``$z$-velocity''
expressed in terms of the momentum for a particle of ``mass'' $E$.
Remarkably, eq.(\ref{eq1}) can
be seen to be equivalent to the classical Hamilton
equations \cite{arnaud} by identifying
$A(z)\equiv iP(z)/X(z)$ where $P(z)$ and $X(z)$ are generalized
(complex) momenta and positions which
satisfy the eqs.(\ref{eq2}). 

Now, we impose an initial condition
at $z=-L$ that the particle has been
prepared into a transverse Gaussian wave-packet,  
specified to have a 
pure real width $\delta_0 $ given by:
\be
\label{eq7}
A_0 = Re[A(-L)] = 1 /(\delta_0)^2;  \qquad  Im[A(-L)]=0
\ee
We assume that the centroid of the initial wave-packet is moving
parallel to the $z$-axis, thus $p_\perp=0$, and $x_\perp=x_0$ is
initially arbitrary at $z=-L$.

The wave-packet enters the lens at 
$z=-\delta z$ and exits at $z=0$. 
Upon entry of the lens $A(-\delta z)$ is given by
the free drift solution
of eq.(\ref{eq1}) from $z=-L$ to the lens, over the drift
distance $L$:
\be
\label{eq8}
A(-\delta z)= \frac{A_0}{(1+i{A_0 L }/{E})}
\ee
where we assume a thin lens, $\delta z/L <<1 $.

In the thin lens, to a good approximation 
for small $\delta z$, we have
from eq.(\ref{eq1}):
\be
\label{eq09}
A(0) = A(-\delta z) + i\delta z K_0
\ee
Here we neglected 
the term $i\delta z A^2/E$ in the differential equation
for $A(z)$ which only
gives negligible free particle spreading in the lens.
Moreover, the classical centroid motion of
the wave-packet is found from eq.(\ref{eq2}):
\be
p_\perp(0) = -K_0 x_0 \delta z; \qquad x_\perp(0) = x_0.
\ee
Upon exiting the lens the particle propagates 
again in
free space a distance $\ell $ with $K=0$. Hence
we find:
\be
\label{eq11}
{A(\ell )} = \frac{A(0)}{1+i{A(0)\ell}/{E}}
\ee
and:
\be
p_\perp(\ell) = -K_0 x_0 \delta z; \qquad x_\perp(\ell) 
= x_0-p_\perp \ell/E.
\ee
Note that the classical trajectory of the
off-axis particle is deflected back toward the
lens axis, $x_\perp=0$.

The focal length, $f$ is defined such that 
$x_\perp(f) =0$, 
hence:
\be
f = \frac{E}{K_0\delta z}
\ee
The kernel of the wave-packet can now be obtained by solving 
eqs.(\ref{eq7},\ref{eq8},\ref{eq09},\ref{eq11}) recursively to obtain:
\be
A(\ell) = \left[
\frac{1-i [L/(E(\delta_0)^2) -(\delta_0)^2 E/f  ]}
{(\delta_0)^2(1-\ell/f) + i({\ell}/{E}+{L}/{E}-
{L\ell}/{Ef})
}
\right]  
\ee

\begin{figure}[t]
\vspace{4cm}
\includegraphics{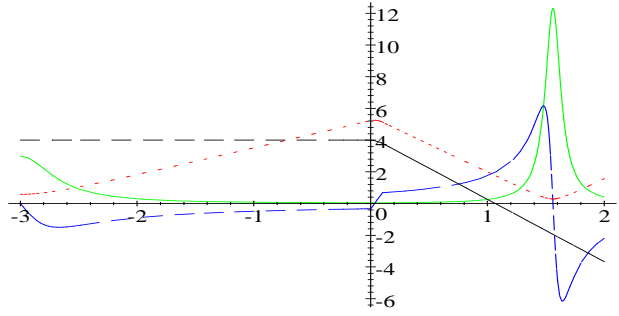}
\vspace{1cm}
\caption[]{
The numerical evolution of a Gaussian wave-function with $L=3.0 $ 
evolving through a lens;  $E=1.0 $, $K_0=10.0 $, $\delta z = 0.1 $
and $A(-3.0)=3.0$, hence $\delta_0=0.577$, and the 
geometrical focal length, $ f=1 $. 
(a) dashed (black) line is $ x_\perp(z) $ crossing 
the axis at the geometrical
focal point; 
(b) dotted (red) Gaussian width $1/\sqrt{Re(A(z))} $ which focuses
at $ f_q $; Also shown are $Re(A(z)) $ (green solid) and $Im(A(z)) $
(blue dot-dashed); note the $Im(A(z)) $ receiving a positive kick
upon passing through the lens. }
\end{figure}

Note that the Gaussian kernel has an imaginary part which changes from
positive (focusing) to negative (defocusing) upon passage of the 
geometrical focal
length, $L>f$.
Thus, the transverse probability distribution becomes:
\be  
|\hat{\phi}|^2 =
{\cal{N}}(z)\exp \left\{ - \left( \frac{(x-x_\perp)^2 }{(\delta_0)^2(1-\ell/f)^2
+ (\ell/E\delta_0 )^2[1-L(\ell-f)/f\ell]^2} 
\right)  \right\}
\ee 
and, the transverse size of the wave-packet is given by:
\be
\delta^2 (\ell) = (\delta_0)^2(1-\ell/f)^2
+ (\ell/E\delta_0 )^2[1-L(\ell-f)/f\ell]^2 
\ee

In Figure 1 we give a numerical integration of the 
Schroedinger equation in the preceding discussion,
which confirms the validity of our solution.  Note that
for finite $L$ the Gaussian width is focused to
a minimum  at $z = f + f^2/L +...$. 
In the limit $L\rightarrow \infty$  the transverse size 
of the wave-packet reaches a minimum 
at the focal point $\ell =f$, where the new effective transverse size is:
\be
\label{eq18} 
\Delta x = \frac{f}{E \delta_0}
\ee
This is the  usual Rayleigh diffractive minimum, $f\lambda/a$
if we  regard $a \sim \delta_0$
as an ``effective aperture size'' through
which the beam has passed, and $E=p_zc = \hbar c/\lambda$,
the usual quantum wavelength of the particle. 

What if the initial prepared wave-function is not the
groundstate of a SHO (pure Gaussian), but is rather an excited
eigenstate of principle quantum number $n$?  Hence, at  $z=-L$,
neglecting $x_\perp$ and $p_\perp$,
we assume:
\be
\label{eq19}
\hat{\psi}(-L) = H_n(x/\sqrt{2}\delta_0) \exp\left( -\frac{1}{2}(x/\delta_0)^2 + C \right)
\ee
where $H_n(\xi)$ is the $n$th Hermite polynomial.

First, we note that the Gaussian solution eq.(\ref{eq4})
contains the generating
function for Hermite polynomials \cite{schiff}:
\be
\label{xeq19}
\hat{\phi} = \exp\left( -\frac{1}{2}A(z)(x)^2 +ip_\perp(z) x + C  \right)
\left( \sum_{n=0}^\infty 
\frac{H_n(\sqrt{ (A(z)/2)}x )}{n!}[ \sqrt{ (A(z)/2)} x_\perp(z) ]^n
\right)
\ee
For a freely drifting particle,
if we choose $p_\perp(-L) =0$, and initial
$x_\perp(-L)=x_0$, in $\hat{\phi}$,
then we see that our solution $\hat{\psi}(z)$ is
determined for any $z$:
\be
\label{xeq20}
\hat{\psi}(z) =  \frac{\partial^n}{\partial x_0^n} \phi (z)|_{x_0=0}
\ee
After passing through an arbitrary lens system, 
the solution for $\hat{\psi}(z)$ becomes messy, and
in general $x_\perp(z)$
and $p_\perp(z)$ are arbitrary, and we cannot
so easily differentiate with respect to $x_0$ to pull
out our solution. 
However, both $x_\perp(z)$
and $p_\perp(z)$ are proportional
to $x_0$ by the linearity of the Hamilton equations.  
At a focal
point we have $x_\perp(f)=0$ (for any $x_0$, owing to linearity), 
and $p_\perp(f) \propto x_0$. 
Hence, the solution
at a focal point simplifies:
\begin{eqnarray}
\label{xeq21}
\hat{\psi}(f) & = & \frac{\partial^n}{\partial x_0^n} \phi (f)|_{x_0=0}
\nonumber \\
& = & \frac{\partial^n}{\partial x_0^n} 
\exp\left( -\frac{1}{2}A(z)(x)^2 +ip_\perp(z) x + C  \right) |_{x_0=0}
\nonumber \\
& \propto  &  x^n \exp\left( -\frac{1}{2}A(z)(x)^2 \right)
\end{eqnarray}
and thus, the arbitrary solution is focused to a Gaussian
times a power of $x$.
This gives a focal spot size:
\be
\label{eq20}
\Delta x = \frac{\sqrt{n} f}{E \delta_0}
\ee
Now this result may seem counterintuitive; we are starting with a
broader initial distribution by the factor $\sqrt{n}$, and we might
guess that this would produce a smaller focal point by an amout
$1/\sqrt{n}$.  The wave-function, however, is not smooth in $x$,
i.e., the Hermite polynomial yields a distribution
of transverse momentum,
and the initial state has ``ears'', 
each of typical Gaussian width $\delta_0$, but
displaced off the optical axis by $\sqrt{n}$ .  These produce
the $\sqrt{n}$ enhancement of the focal spot.  Yet another way to
see this is to note that one can make a classical off--axis 
centroid motion of the groundstate Gaussian
by superimposing large $n$ states, and
the Gaussian width will yield the minimal $f/\delta_0E$ result.
Of course, the quantum state of interest
to us will typically have a large value of $n$ determined
by radiative relaxation. We consider this in the next section.

The actual linear acceleration phase is inconsequential
to this result.
The above discussion assumed a uniform drift 
in the longitudinal $z$-direction, i.e., constant energy $E$.
If the particle is accelerating linearly, then $E$ becomes
 z-dependent, $E(z)= (E_f-E_0)z/L + E_0$.
It is easily seen that the only effect on our solution
is to replace $L$ by $L\ln((E_f-E_0)/E_0)$, where $E_0$ is
the initial energy, $E_f$ the final energy,
and $E$ in the above expressions
is everywhere replaced by $E_f$.
For the first NLC, we have $E_f\sim 250$ GeV $ >> E_0\sim 2$ GeV.
The linear acceleration phase
is thus equivalent to free drift through an effective distance
of $L\ln(E_f/E_0) \sim 45$ km where $L=10$ km. 
The amount by which a wave-packet of initial size of
$25$ nm spreads throughout
the NLC acceleration phase is about a factor of $6$.
However, this spreading is irrelevant to computing the final diffractive
limit as seen in eq.(\ref{eq18}) where
the free drift length $L$ has completely cancelled
from the expression at the classical focal point, and
only the initial quantity $\delta_0$ 
(together with the local quantities $f$ and $E$)
controls the diffractive 
limit.\footnote{We remark that the proper way to view the quantum
spreading in the transverse phase-space is to use Wigner functions, which
depend upon both $x$ and a quantum momentum $p$.  The Wigner function
isocontours deform in a manner that is conformal to the classical emittance
envelope, so while the wave-functions spread in $x$ the 
Wigner envelopes  actually shear
in $x$ and $p$ and remain contained in the transverse phase-space.\cite{hill} 
}

Thus, the ultimate diffractive limit is controlled by
the initial boundary conditions on the wave-function size,
i.e. by $\delta_0$, and not by the intervening unitary lens system.
What 
in general determines $\delta_0$? For the NLC the initial
wave-function, as well as
the initial classical distribution, is prepared in the ``damping rings.''
Damping rings are essentially a system of magnets arranged
as wigglers which induce synchrotron radiation and cool
the classical beam bunches of electrons. They are designed to produce
roughly a four order of magnitude reduction in one of the transverse
dimension phase space volumes, i.e., $\sim \Delta x \Delta p_x$ 
(the transverse emittance). As the system cools classically, it is
also relaxing quantum mechanically. This occurs because the particles
in the wiggler chain experience a transverse SHO potential, and
synchrotron radiation pushes highly excited wave-functions toward the 
Gaussian groundstate in this potential \cite{huang}. However,
there are also re-excitation transitions which eventually come
into equilibrium, and a typical average SHO principle quantum
number is established. 
While this is certainly an
oversimplified view of the actual system, we will 
use it as a starting point
to estimate $\delta_0$.

\section{Magnetic Focusing and Damping}

We now sumarize
the details of the motion of a
transverse wave-packet in a magnetic field.
This is discussed in detail in the classic work
of Sokolov and Ternov \cite{sokolov}.  We will use
the more transparent WKB approximation, 
expanding about the classical radius of motion.
Hence, one should use caution in comparing solutions,
e.g., principle quantum numbers refer to different things.
For example, large $n$ in the usual framework \cite{sokolov}
corresponds to small $n$, but large classical radius $j_z$
presently.

Consider a particle moving in
a planar orbit in a uniform magnetic field,
aligned in the $\hat{z}$
direction, $\vec{B} = \hat{z} B_0$
in a cylindrical
coordinate system $(r, \phi, z)$.  The vector potential  
can be chosen as $A_\phi = rB_0/2$
with $A_r=0$ and $A_z=0$. 
We examine the transverse motion of a relativistic
electron in the plane $z=0$. For an anzatz of the form 
$e^{-iEt/\hbar}\psi(r,\phi)$, the KG equation
becomes:
\be
\label{eq9}
 \left[ -E^2 +m^2 -\frac{\partial^2 }{\partial r^2}
- \frac{1}{r}\frac{\partial }{\partial r }
+ \left(\frac{i}{r}\frac{\partial}{\partial \phi}-eA_\phi
\right)^2
\right]\psi =0
\ee
where a possible momentum
component $p_z$ has been set to zero. 
Consider a state of with a
large ``pseudo-angular momentum,'' $\ell$, 
and scale out a factor
of $1/\sqrt{r}$:
\be
\psi = \frac{1}{\sqrt{r}}e^{i\ell\phi} \hat{\chi}(r,t).
\ee
($\ell$ is not  
the physical angular momentum because it is gauge dependent
due to presence of  the vector potential; the physical angular
momentum in the present case is $2\ell$, as we will see below).
Hence:
\be
\left[-E^2 +m^2 -\frac{\partial^2 }{\partial r^2}
+ \left(-\frac{\ell}{r}- \frac{1}{2}erB_0
\right)^2 +\frac{1}{4r^2}
\right]\hat{\chi} = 0
\ee 
This now has the apparent form
of a one-dimensional Schroedinger 
equation with an effective potential:
\be
V(r) = \left[ \left(\frac{\ell}{r}+\frac{1}{2}erB_0
\right)^2+\frac{1}{4r^2}\right]
\ee
The potential has a minimum at:
\be
\label{eq29}
r = R_0 \equiv \left( \frac{\sqrt{4\ell^2+1}}{eB_0} \right)^{1/2}\rightarrow 
\sqrt{\frac{2\ell}{eB_0}}
\ee
where the latter expression corresponds to
$\ell >> 1$. Consider henceforth $\ell >>1.$
We consider small radial
fluctuations around the large orbital
radius $R_0$ as $ r = R_0 + x$ and expand:
\be
\left[ -E^2 +m^2 -\frac{\partial^2 }{\partial x^2}
+ V(R_0) + \frac{1}{2} x^2 V''(R_0)
\right]\hat{\chi}=0
\ee 
and:
\be
V(R_0)= {2eB_0\ell}   \qquad V''(R_0)= 2{e^2B_0^2}
\ee
Thus the high orbital angular momentum Landau levels
are approximate eigenstates of the SHO potential defined
by $\hat{V}(x) = \frac{1}{2}x^2 V''(R_0)$, or $K=e^2B_0^2/2E$.
The states are labelled by $(n, \ell)$ where $n$ is a
principle SHO quantum number;
the energy eigenvalues of these levels are given by:
\be
{{E}^2} = m^2 +  {eB_0}(2\ell +n +\frac{1}{2} ) 
\ee
In the presence of the gauge interaction the physical
angular momentum is  $L_z= -i\partial/\partial \phi + eR_0 A_\phi(R_0) $.  Hence, 
the physical angular
momentum is:
\be
j_z = \ell + \frac{1}{2}eB_0R_0^2 = 2\ell
\ee
where we use the explicit solution for $R_0$ from eq.(\ref{eq29}) in the
latter expression.  Therefore, to make the classical
correspondence, we identify
the angular momentum with that of an entering beam particle of
momentum $p_\phi$, to obtain $R_0p_\phi = j = 2\ell$.  This yields
consistency with the familiar expression
for the classical orbital radius and the total energy:
\be
R_0 = \frac{p_\phi}{eB_0} \qquad E^2 = m^2 + {eB_0}(j_z + n + \frac{1}{2} ) 
\ee
  Transitions that increase  $n$, but decrease
$j_z$ are allowed; hence synchrotron radiation can be
excitatory as well as relaxational. The fact that 
the energy is degenerate,
depending upon the combination $j_z + n$ is a consequence
of the symmetry in the choice of the classical orbital center.
(Note that the solution formed with the anzatz $e^{-i\ell \phi}$
for large $\ell$ is actually a solution of vanishing 
physical momentum; it is a zero-mode associated
with the translational invariance of the center of
the particle's orbit).

The groundstate in the transverse dimension is a Gaussian 
with $A= |eB_0| $, given by:
\be
\delta_0 = \frac{1}{\sqrt{Re(A)}} = (eB_0/\hbar c)^{-1/2}. 
\ee
For a typical field strenth of $1$ Tesla we obtain $\delta_0
\sim 25$ nm. 
The ``spring constant'' is $O(e^2)$, hence we say that the 
dipole magnet is weakly focusing (for quadrupoles 
$V''(R_0)\sim {eB_0}E/a$, where $a$ defines
a gradient, hence ``strong focusing''). 
This description applies
to wigglers, even though the dipole magnet field is alternating
in $z$, if the magnitude of the $B$ field is roughly constant.

It has been known for a 
long time that an equilibrium between deexcitatory
and excitatory transitions for a particle in a damping system
(or synchrotron)
will be established \cite{sands}, and there will be an equilibrium value of
$n$.  This value is roughly estimated as follows.
The typical energy of synchrotron radiated photons is \cite{sands},
\cite{sokolov}: 
\be
E_\gamma \sim \frac{1}{R_0}\left( \frac{E}{m_e} \right)^3
\ee
A unit step in a quantum number $n$ or $j_z$ produces
only a small energy change, $\sim eB_0/E \sim 1/R_0$.
The dipole approximation selection
rules imply large allowed changes in $j_z$, but
only unit steps in $n$:
\be
\Delta j_z \sim \frac{E}{R_0eB_0}\left( \frac{E}{m_e} \right)^3 
\sim \left( \frac{E}{m_e} \right)^3
\qquad \qquad
\Delta n \sim \pm 1
\ee
Bear in mind that we are treating $n$ as the principle quantum number in the
WKB focusing channel defined by expanding about $R_0$, and dipole transitions
involving the operator
$\vec{A}\cdot \nabla $ will change $n$ by a unit (these can 
be excitatory).
Then $\Delta j_z/R_0$ is essentially the change
in longitudinal electron momentum, imparted to the photon. In transitions,
though $R_0$ changes, there is no sudden translation in the transverse
position of the electron wave-function, only a transition in motion,
i.e., the virtual
center of the orbit changes \cite{sokolov, huang}.

Over a radiative energy loss time interval 
the number of emitted photons is:
\be
n_\gamma \sim \frac{E}{E_\gamma} \sim \frac{m^3_e}{eB_0 E }
\sim {R_0}\left( \frac{m^3_e}{E^2 }\right)
\ee
The principle quantum number $n$
undergoes a random walk by roughly $\sqrt{n_\gamma}$, 
hence the  equilibrium $\bar{n}$ is of order $\sim \sqrt{n_\gamma}$.
Using $E\sim 2$ GeV, and $B_0 \sim 1 $ Tesla, whence $R_0 \sim 6.6$ m,
we find $n_\gamma = 1.12\times 10^{6}$ and
$\bar{n} \sim 1.06\times 10^3$.

Hence, our diffractive limit is now increased by $\sqrt{\bar{n}}
\sim 0.33\times 10^2$,
and we thus have a beam spot size $\sqrt{\bar{n}}\times 0.06
\sim 2.0$ nm.  Why is this result so close to the design
goals of the NLC that are obtained by classical physics?
Indeed, we believe that this result {\em is} a quantum derivation
of the classical result!  The quantum number $n$ scales
as $1/\hbar$, while our diffractive limit scales as 
$\Delta x \propto \sqrt{\hbar}$,
hence the product $\sqrt{n}\Delta x$ is independent of $\hbar$.
This, moreover, assures us that the ultimate quantum limit
is of order $1/\sqrt{\bar{n}}$ smaller than the minimal
classical analysis.  (We note that the Oide effect \cite{chen}
may be understood as a blowing up of $\bar{n}$ in intense final
focus magnets, where large transverse energy photons are radiated).

A more detailed discussion of synchrotron 
radiation relaxation is beyond
the scope of the present paper. Excellent
treatments  can be found in \cite{sokolov}, \cite{huang2},
and the pioneering work of \cite{sands}.

\section{Quantum Particle in a Synchrotron} 

The solution to the Schroedinger
equation for passage of
a free particle through a lens, eq.(\ref{eq0}), 
can be completely described by the  simple 
classical optical matrix
methods. 
If one passes a classical ray moving in the
$\hat{z}$ direction through a lens system,
the outgoing state of the transverse
$\hat{x}$ canonical variables may be
written as \cite{edwards}:
\be
\left( \begin{array}{c}
x(\ell) \\
p(\ell)/E
\end{array}\right)_{out} = {\cal{M}}
 \left( \begin{array}{c}
x(-\delta z) \\
p(-\delta z)/E
\end{array}\right)_{in} \qquad \makebox{where:}\;\;\; \det{\cal{M}}= 1.
\ee
The unimodular matrix ${\cal{M}}$ for a compound
sequence of lens elements is the corresponding sequential product
of the individual matrices of the elements.
For example, a sequence of free propagation
(distance $L$), followed by defocusing lens
(focal length $-f$), followed by a space 
$a$, followed by a focusing lens (focal length $f$),
followed by free propagation
(distance $\ell$) yields the result:
\be
\label{eq32}
\left( \begin{array}{c}
x(\ell) \\
p(\ell)/E
\end{array}\right) = \left( \begin{array}{cc}
 1+\frac{a}{f}-\frac{a\ell}{f^2} & \;\;(L+a+\ell)+ 
 \frac{aL}{f}- \frac{a\ell}{f}- \frac{aL\ell}{f^2}  \\
 -\frac{a}{f^2} & 1-\frac{a}{f}-\frac{aL}{f^2}
\end{array}\right) 
 \left( \begin{array}{c}
x(-\delta z) \\
p(-\delta z)/E
\end{array}\right) 
\ee
The zero
of the $(11)$ matrix element in $\ell = F\equiv f+f^2/a$
implies the system is net focusing with composite focal
length $F$ (e.g., see ref. \cite{edwards}).

The effect of this particular 
lens system in quantum mechanics,
e.g., on the
Gaussian kernel $A$ as defined in eq.(\ref{eq4}), can be easily derived
from the Schroedinger equation:
\be
\label{eq33}
A(\ell) = \frac{A_0(1-a/f-aL/f^2) + i Ea/f^2}{1 + a/f -a\ell/f^2 + 
         i A_0 /E [L+a+\ell +aL/f -a\ell/f - (a\ell L/f^2)]}
\ee
The focal length, $F$, is where ${\cal{M}}_{11} =0$,
and, at the focal length we obtain the width:
\be
\delta(F) = \frac{ f^2}{aE\delta_0 } \qquad \makebox{where:}\;\;
Re(A(0))=\frac{1}{\delta^2_0 }
\ee
This result is the minimal diffractive quantum limit for the
composite lens system, and it  is again determined by the initial
width of the quantum state.\footnote{Here we might imagine taking $a\rightarrow
\infty$ holding $f$ fixed to cause $\delta(F)\rightarrow 0$; however, for
$a >> f$ the longitudinal dimension ($\Delta z$) of the 
focal point becomes small as $\sim f/a$
owing to the $aL/f$ and $a\ell L/f^2$ terms in ${\cal{M}}_{12}$, and
the finite longitudinal distribution of the beams 
becomes problematic;
we have not looked in detail at optimization of this.}

Of course, beyond $\delta_0$, there is actually 
no new information in the above formula for $A$
than is already present in the lens matrix for the classical ray
optics. If the lens matrix is ${\cal{M}}_{ij}$, then we see
by comparison with eq.(\ref{eq33})
the general result for the Gaussian kernel (\cite{kogelnik}, \cite{arnaud}):
\be
\makebox{{\bf Theorem I:}}\qquad
A_{out} = \frac{{\cal{M}}_{22}A_{in} - i{\cal{M}}_{21}E }{ {\cal{M}}_{11}+ 
         i{\cal{M}}_{12}A_{in} /E }
\ee
That is, if we write the out-amplitude as:
\be
A_{out} = -i E\left(\frac{{\cal{N}}}{{\cal{D}}}\right)
\ee
then:
\be
\left( \begin{array}{c}
{\cal{D}} \\
{\cal{N}}
\end{array}\right) = \left( \begin{array}{cc}
{\cal{M}}_{11} & {\cal{M}}_{12}\\
{\cal{M}}_{21} & {\cal{M}}_{22}
\end{array}\right) 
 \left( \begin{array}{c}
-i E \\
A_{in}
\end{array}\right) 
\ee
From this result we can easily derive the quantum limit on
beam size at the classical focal length.  The classical focal
length occurs where ${\cal{M}}_{11}=0$.  At this point we have
${\cal{M}}_{12}=-{\cal{M}}^{-1}_{21}$.  Hence we readily find:
\be
\makebox{{\bf Theorem II:}}\qquad
Re A_f = \frac{E^2 Re A_0}{{\cal{M}}^2_{12}[(Re A_0)^2 + (Im A_0)^2]}
= \frac{1}{(\Delta x)^2}.
\ee
If $A_0=\hbar/\delta_0^2$ is pure real, then:
\be
\Delta x =  \frac{{\cal{M}}_{12}}{E\delta_0} \equiv
 \frac{ \tilde{F}}{E\delta_0}
\ee 
noting that ${\cal{M}}_{12}\equiv \tilde{F} \sim f
$ is a length scale comparable
to the focal length at the classical focal length, e.g.,
$\tilde{F} = f^2/a$ in our previous compound lens example
(see previous footnote).

Now, if the magnet system is periodic, as in a synchrotron, we 
expect quantum states that are approximately 
periodic solutions in the matrix.  
Periodic solutions must be
eigenstates of the matrix ${\cal{M}}$.  
Consider first the motion within a very thick lens, 
i.e., a continuous transverse
SHO potential. 
For an infinitesimal displacement in the $z$-direction,
the lens matrix is:
\be
{\cal{M}}_{SHO} = \left( \begin{array}{cc}
1 & \delta z\\
-\frac{K\delta z}{E} & 1
\end{array}\right) 
\ee
which is unimodular to ${\cal{O}}((\delta z)^2)$. The eigenvalues
of ${\cal{M}}_h  $ are $\lambda_\pm = 1 \pm i\delta z \sqrt{K/E}$.  The
stable, quantum solutions in the lens
are therefore the eigenvectors:
\be
\lambda_\pm \left( \begin{array}{c}
-iE \\
A_0
\end{array}\right) = \left( \begin{array}{cc}
{\cal{M}}_{11} & {\cal{M}}_{12}\\
{\cal{M}}_{21} & {\cal{M}}_{22}
\end{array}\right) 
 \left( \begin{array}{c}
-i E \\
A_{0}
\end{array}\right) 
\ee
hence we find:
\be
A_0 = \pm \sqrt{KE} 
\ee
Hence, the stable solution in
the linear focusing channel is, indeed, the Gaussian groundstate 
solution in the SHO potential.  We have simply recovered the
usual Gaussian groundstate in this limit.

Consider now a ``synchrotron,'' i.e., 
periodic magnet lens lattice based upon the 
above lens configuration.  We assume
an infinite series of alternating dipoles with spacing $a$ 
and focal lengths $\pm f$.  Replacing $L=a-\ell$ in
the matrix elements ${\cal{M}}_{ij}$ of
eq.(\ref{eq32}) gives the lens matrix for the synchrotron:
\be
{\cal{M}} =
\left( \begin{array}{cc}
 1+\frac{a}{f}-\frac{a\ell}{f^2} & \;\;2a+ 
 \frac{a^2}{f}- 2\frac{a\ell}{f}- \frac{a^2\ell}{f^2}+\frac{a\ell^2}{f^2}  \\
 -\frac{a}{f^2} & 1-\frac{a}{f}-\frac{a^2}{f^2}+\frac{a\ell}{f^2}
\end{array}\right) 
\ee
The condition that we have a periodic solution
is: 
\be
A_{\ell} = \frac{{\cal{M}}_{22}A_{\ell } - i{\cal{M}}_{21}E }{ {\cal{M}}_{11}+ 
         i{\cal{M}}_{12}A_{\ell } /E }
\ee
Using $\det{\cal{M}}=1$ we find:
\be
A_{\ell} = \frac{-i E}{2{\cal{M}}_{12}}\left[{\cal{M}}_{22}- {\cal{M}}_{11} \pm
\left( ({\cal{M}}_{22}+ {\cal{M}}_{11})^2 -4 \right)^{1/2}\right]
\ee
Stable quantum solutions (solutions that are normalizeable Gaussians
for all $\ell$) therefore require:
\be
-1 \leq \frac{1}{2}Tr({\cal{M}}) \leq 1
\ee
This is, of course, the familiar stability condition for
the classical motion.
Note that $Tr({\cal{M}}) = 2- a^2/f^2$, which is $\ell$ independent,
thus when the condition is met for particular choices of $f$ and $a$
it holds everywhere.  The stability condition is the usual one, 
$f\geq a/2$.

The solution for $A(\ell)$ is:
\be
A(\ell) = \frac{E}{[(2f+a)(1-\ell/f)+\ell^2/f ] }
\left[ \left( 1-\frac{a^2}{4f^2} \right)^{1/2} + i \left( 1 +\frac{a}{2f} -
\frac{\ell}{f} \right)  \right]
\ee
where $f\geq a/2$ and $0\leq \ell \leq a$.

Consider the special case of a system in which $f=a$. We see
that the minimum Gaussian width occurs at $\ell = a$, given by:
\be
min(1/Re{A(\ell)}) = \frac{2 f}{\sqrt{3}E} \equiv (\Delta x)^2
\ee
This implies that the minimum 
achievable beam spot
size in a synchrotron is $\sim \sqrt{f/E}$. 
If a focusing magnet of focal length $f'$ is inserted into
the synchrotron magnet lattice and, then we obtain
the minimum spot size $\sim f'/E \Delta x \sim f'/\sqrt{E f} $.
There is no initial parameter $\delta_0$ since we have assumed that
synchrotron radiation relaxes the quantum state into the stable,
periodic solution. Here we see a
potential  advantage of a linear collider over a synchrotron, in
that the linear collider has a much larger $\delta_0\sim \sqrt{f/E_0}$
prepared in the low energy damping ring, which makes the quantum 
diffractive limit 
smaller, $\sim f/E\delta_0$, while in 
the synchrotron $\delta_0\sim \sqrt{f/E}$, where $E$ is the
larger beam energy, thus giving a larger diffractive
limit $\sim \sqrt{f/E}$.  

\section{Conclusion}
 
In conclusion,
we estimate  the minimal quantum beam spot size 
achievable in a 
linear collider
to be given by:
\be
\Delta x \geq \frac{\hbar c f}{E\delta_0}
\ee
where $f$ is the final focal length, $E$ the beam energy, and
$\delta_0$ is the initial transverse size of the wave-functions
prior to acceleration. This may be
viewed as a direct transcription of the Heisenberg
uncertainty principle.   $\delta_0$ is prepared in the synchronous 
damping rings, typically wigglers, and $\delta_0 \sim 1/\sqrt{eB/\hbar c}$
where $B$ is the magnetic field strength.  We have for an NLC-class
machine,
$B\sim 1$ Tesla, $f\sim 2 $ m, $E\sim 250$ GeV 
hence $\delta_0 \sim 25$ nm, and $\Delta x
\sim {\cal{O}}(0.06)$ nm.  

Radiation damping implies that the initial state wave-function is not
a groundstate, and has an average equilibrium principle quantum number $
\bar{n}$.
Then our result is modified:
\be
\Delta x = \sqrt{\bar{n}}\frac{\hbar c f}{E\delta_0}
\ee
$\bar{n}$ is estimated to be
\be
\label{eq459}
\bar{n} \sim \left(\frac{m^3_e}{eB_0 E } \right)^{1/2}
\ee 
or, for the above parameters, $\bar{n} \sim 0.11 \times  10^4$. 
Eq.(\ref{eq459}) is essentially classical, and yields $\Delta x \sim 2$ nm,
roughly consistent with the classical 
vertical final focus beam spot size
of the NLC,  
$\sim 5$ nm.  A more precise analysis of this latter
effect is, however, certainly required. This may be a useful way to 
approach other phenomena, such as the Oide effect, 
in which large fluctuation in
$\bar{n}$ in strong final focus magnets can occur,
broadening the beam spot size.
 
We have examined the quantum solution
in a simple FODO synchrotron model. 
In a synchrotron information about
the initial $\delta_0$ is lost,
and the minimal transverse beam spot size is:
\be
\Delta x = \sqrt{\frac{\hbar c f}{E}}
\ee
which is ${\cal{O}}(1)$ nm for most high energy
synchrotrons, e.g., LEP and Tevatron, in operation
at present. Again, a factor of $ \bar{n} $ would
yiedl the classical result. Presumably proton synchrotrons
are far from the equilibrium, with  $ n >> \bar{n} $.

A tantalizing question is: can quantum diffractive 
effects be observed?  More generally, our discussion
has been motivated by the belief that
quantum optics may be the preferred way to analyze
futuristic machines. A more general formalism more
symmetrical in $p_\perp$ and $x_\perp$ for the
study of the quantum phase space, perhaps
based upon Wigner's formulation of quantum mechanics,
is desired \cite{hill}.

 \vspace*{1.0cm}
 \noindent
 {\large \bf Acknowledgements}
 
 \noindent
 We wish to thank W. Bardeen, D. Burke, J.~D.~Jackson,
 R. Noble, C.Quigg,
 A. Tollestrup, and especially P. Chen, D. Finley, L. Michelotti,
 and R. Raja for useful discussions.
\newpage


\vfil
\end{document}